\documentclass[]{aastex631}
\usepackage{CJK}
\usepackage{amssymb}
\usepackage{amsmath}
\usepackage{algorithm} 
\usepackage{algpseudocode} 
\usepackage{makecell}
\shorttitle{$\alpha$-Deep Probabilistic Inference ($\alpha$-DPI)}
\shortauthors{He Sun et al.}
\graphicspath{{./}{figures/}}

\begin{document}
\begin{CJK*}{UTF8}{gbsn}
\title{$\alpha$-Deep Probabilistic Inference ($\alpha$-DPI): efficient uncertainty quantification from exoplanet astrometry to black hole feature extraction}

\author[0000-0003-1526-6787]{He Sun}
\affiliation{California Institute of Technology, 
1200 East California Boulevard, 
Pasadena, CA 91125, USA}

\author{Katherine L. Bouman}
\affiliation{California Institute of Technology, 
1200 East California Boulevard, 
Pasadena, CA 91125, USA}

\author{Paul Tiede}
\affiliation{ Perimeter Institute for Theoretical Physics, 31 Caroline Street North, Waterloo, ON, N2L 2Y5, Canada}
\affiliation{ Department of Physics and Astronomy, University of Waterloo, 200 University Avenue West, Waterloo, ON, N2L 3G1, Canada}
\affiliation{ Waterloo Centre for Astrophysics, University of Waterloo, Waterloo, ON N2L 3G1 Canada}
\affiliation{ Black Hole Initiative at Harvard University, 20 Garden Street, Cambridge, MA 02138, USA}
\affiliation{ Center for Astrophysics | Harvard \& Smithsonian, 60 Garden Street, Cambridge, MA 02138, USA}

\author{Jason J. Wang}
\affiliation{California Institute of Technology, 
1200 East California Boulevard, 
Pasadena, CA 91125, USA}

\author{Sarah Blunt}
\affiliation{California Institute of Technology, 
1200 East California Boulevard, 
Pasadena, CA 91125, USA}

\author{Dimitri Mawet}
\affiliation{California Institute of Technology, 
1200 East California Boulevard, 
Pasadena, CA 91125, USA}



\begin{abstract}


Inference is crucial in modern astronomical research, where hidden astrophysical features and patterns are often estimated from indirect and noisy measurements. Inferring the posterior of hidden features, conditioned on the observed measurements, is essential for understanding the uncertainty of results and downstream scientific interpretations.
Traditional approaches for posterior estimation include sampling-based methods and variational inference. However, sampling-based methods are typically slow for high-dimensional inverse problems, while variational inference often lacks estimation accuracy. In this paper, we propose $\alpha$-DPI, a deep learning framework that first learns an approximate posterior using $\alpha$-divergence variational inference paired with a generative neural network, and then produces more accurate posterior samples through importance re-weighting of the network samples. It inherits strengths from both sampling and variational inference methods: it is fast, accurate, and scalable to high-dimensional problems. We apply our approach to two high-impact astronomical inference problems using real data: exoplanet astrometry and black hole feature extraction.

\end{abstract}


\section*{} \label{sec:intro}

Inferring hidden features from indirect, sparse and noisy observational data is a fundamental challenge in modern astronomical research. When tackling these inverse problems, it is important to consider uncertainty in the inferred solution via a posterior rather than simply a point estimate (e.g., Maximum {\it a posteriori}, MAP). A full posterior guarantees that all possible scientific interpretations have been considered. However, due to the large amount of observational data, the high dimensionality of features to infer, and the potential multi-modality of  posterior distributions, recovering a full posterior distribution is often computationally challenging. In this paper we propose an efficient and flexible approach for full posterior estimation in inverse problems, and demonstrate the approach on two applications in astronomical inference. 

Traditionally, sampling-based approaches, such as importance sampling and Markov Chain Monte Carlo (MCMC) methods, are widely used to solve inference problems in computational science. Importance sampling first generates random samples based on a proposal distribution and then re-weights the samples according to their likelihood to approximate the true posterior; MCMC gradually refines the samples using a Markov chain transition distribution to create samples from a posterior distribution. Although these methods have achieved great success in many Bayesian inference problems, they typically suffer from the curse of dimensionality in high-dimensional estimation. Without a good proposal distribution in importance sampling or a good random initialization in MCMC, sampling approaches are often prohibitively slow to converge \citep{mcbook}. 

To overcome the above limitation, optimization-based inference approaches, such as variational inference (VI), have drawn increased attention. Instead of directly sampling the parameter space, VI methods introduce a parametric density function (e.g., a Gaussian mixture model), and solve the inference problem by minimizing a loss (e.g., KL-divergence) between the variational density function and the target posterior distribution. VI methods are drastically more efficient than sampling-based approaches since their optimization is gradient-based. However, limited by the modeling capacity of the variational density function and the choice of divergence loss, traditional VI methods often estimate overly-simplified distributions and sometimes lead to degenerated posterior estimation (e.g., mode collapse) \citep{zhang2018advances}.
 
In this paper, we propose a new deep learning Bayesian inference approach, $\alpha$-Deep Probabilistic Inference ($\alpha$-DPI), for fast and accurate posterior estimation. Our proposed approach consists of two primary steps: 1) we first use $\alpha$-divergence VI to optimize a normalizing flow generative model to approximate a posterior, 2) we then use importance sampling to recover a more accurate distribution. 
By utilizing current state-of-the-art deep neural network architectures, $\alpha$-DPI inherits advantages from both sampling and optimization methods.
This proposed new deep learning approach not only significantly improves the computational efficiency and accuracy of high-dimensional posterior estimation, but also naturally leads to a metric that can be used for model selection when the underlying physical model is unknown. 
We apply our method to two difficult high-impact astronomical inference problems: exoplanet astrometry and black hole feature extraction. In both cases, $\alpha$-DPI achieves improved performance in term of estimation efficiency and accuracy when compared with traditional methods.

\section{Results} \label{sec:results}
\subsection{Overview of the method} \label{subsec:overview}
$\alpha$-DPI is a two-step algorithm that combines $\alpha$-divergence VI, a normalizing flow neural network, and importance sampling for uncertainty quantification with Bayesian inference. Using only the observed measurements $y$, the goal of $\alpha$-DPI is to generate samples of the hidden state $x$ that approximate those from the target posterior $p(x|y)$. For a more detailed description of the method please see Section~\ref{sec:methods}.

First, $\alpha$-DPI learns the weights $\theta$ of a normalizing flow neural network $g_{\theta}(\cdot)$ that generates samples from a proposal distribution $q_{\theta}(x)$. This generative network is optimized using an $\alpha$-divergence VI loss:
\begin{equation} \label{eq:method-vi}
    \theta^{\star} = \arg \min_{\theta} D_{\alpha}[q_{\theta}(x)\|p(x|y)] \approx \arg \min_{\theta} \frac{1}{N} \sum_{n=1}^N \left[\exp [(1-\alpha) (\log p(y|x_n) + \log p(x_n) - \log q_{\theta}(x_n))] \right],
\end{equation}
where $D_{\alpha}[A\|B]$ is the Renyi's $\alpha$-divergence of $A$ from $B$ \citep{renyi1961measures, van2014renyi}, $x_n = g_{\theta}(z_n)$ for $z_n \in \mathbb{R}^{|x|} \sim \mathcal{N}(0,1)$ such that $x_n \sim q_{\theta}(x)$, and $p(y|x_n)$ and $p(x_n)$ should be differentiable functions to facilitate gradient-based optimization. 
In this $\alpha$-divergence formulation, $\alpha$ is a selected value between 0 and 1.
Optimizing with an $\alpha$-divergence loss encourages the learned distribution $q_{\theta}(x)$ to be similar to $p(x|y)$.
Refer to Section~\ref{subsubsec:alphadiv} for a more detailed description of $\alpha$-divergence and its relation to Kullback-Leibler 
(KL) divergence.
The resulting generative network will therefore map samples $z_n$ from an {\it i.i.d} Gaussian distribution to those from the learned proposal distribution $q_{\theta}(x)$, where $q_{\theta}(x)$ approximates the target posterior $p(x|y)$. 

Second, $\alpha$-DPI performs importance sampling to re-weight samples from the learned proposal distribution. 
In particular, each sample in the set $\{x_j\}$, where $x_j=g_{\theta}(z_j) \sim q_{\theta}(x)$, is weighted by
\begin{equation} \label{eq:method-is}
    w(x_j) = \frac{p(y|x_j)p(x_j)}{q_{\theta}(x_j)},
\end{equation}
so that $p(x_j|y) \propto w(x_j) q_{\theta}(x_j)$. By simply re-sampling from the set $\{x_j\}$ according to $\{w(x_j)\}$, $\alpha$-DPI produces a new set of samples $\{ x'_j\}$ that better captures $p(x|y)$.


A critical building block of $\alpha$-DPI is the normalizing flow generative neural network. This class of invertible networks is widely used in computer vision and machine learning for density function approximation. 
Normalizing flow generative networks are able to capture complex correlations between parameters in the target posterior. 
Therefore, unlike simple distribution models typically used for VI (e.g., Gaussian), 
these networks lead to better proposal distributions that in turn lead to more efficient importance sampling. Additionally, unlike Variational Autoencoders (VAEs) that rely on efficient optimization via the Gaussian reparameterization trick, normalizing flows are bijective and therefore are not restricted to capturing unimodal distributions. 
However, note that since a traditional normalizing flow architecture will result in a continuous generative distribution, target distributions with disconnected modes cannot be captured by a normalizing flow without a connecting ``bridge."

$\alpha$-DPI significantly outperforms our previous work for image reconstruction uncertainty quantification: deep probabilistic imaging \citep[DPI][]{sun2021deep}.
In \cite{sun2021deep} a single-step KL-divergence VI was used to approximate the target posterior distribution; this formulation would sometimes result in mode collapse. In contrast, we find that formulating the inference using $\alpha$-divergence, which trades-off exploration and exploitation through the $\alpha$ parameter, is far less susceptible to mode collapse. $\alpha$-divergence VI has the disadvantage of generally producing a distribution $q_{\theta}(x)$ that samples $x$'s outside of the target posterior. However, by simply following $\alpha$-divergence VI with importance sampling we can more accurately capture complex multi-modal posterior distributions than KL-divergence VI. 
In this paper, we denote the method presented in \cite{sun2021deep} as KL-DPI.


\subsection{Exoplanet astrometry and orbital fitting with Gemini Planet Imager (GPI)} \label{subsec:exoplanet}
\begin{figure}[!h]
\begin{minipage}[t]{0.4\linewidth}
\vspace{0pt}
\centering
\includegraphics[width=0.99\textwidth]{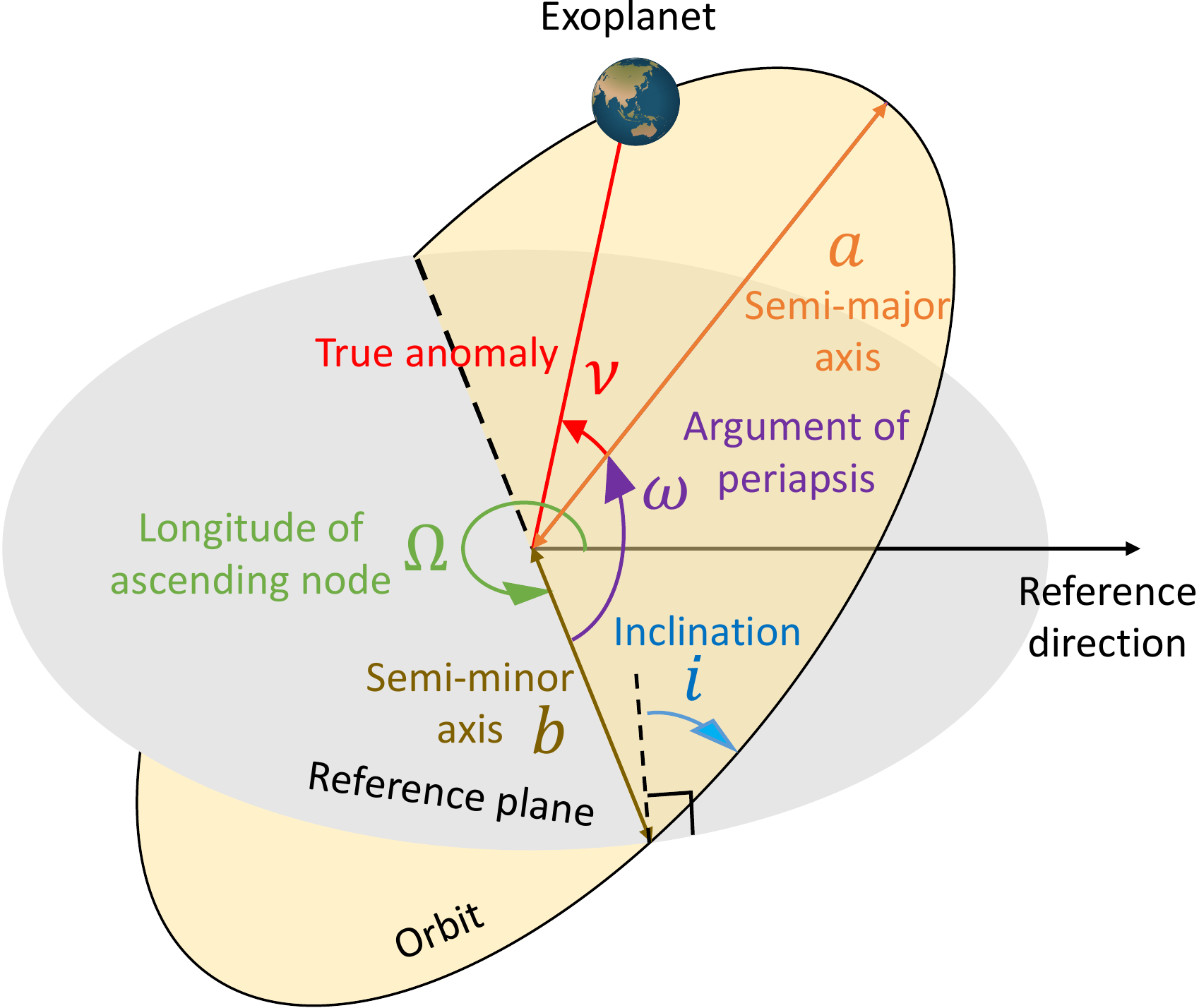}
\end{minipage}\hfill
\begin{minipage}[t]{0.6\linewidth}
\vspace{12pt}
\centering
 \begin{tabular}{c c c} 
 \hline
 Parameter & Unit & Prior distribution\\ [0.5ex] 
 \hline\hline
 semimajor axis ($a$) & astronomical unit ($au$) & log uniform: $\log U(10, 10^4)$\\ 
 eccentricity ($e$) & $-$ & uniform: $U(10^{-8}, 0.99)$\\ 
 inclination angle ($i$) & degree ($\circ$) & uniform: $U(0, 180)$\\ 
 \vtop{\hbox{\ argument of}\hbox{periastron ($\omega$)}} & degree ($\circ$) & uniform: $U(0, 360)$\\ \vtop{\hbox{\quad \ \ longitude of}\hbox{ascending node ($\Omega$)}} & degree ($\circ$) & uniform: $U(0, 360)$\\ 
 \vtop{\hbox{\quad epoch of}\hbox{periastron ($\tau$)}} & $-$ & uniform: $U(0, 1)$\\ 
 parallax ($\pi$) & milliarcsecond ($mas$) & Gaussian: $N(56.95, 0.26)$\\ 
 total mass ($M_T$) & solar mass ($M_{\bigodot}$) & Gaussian: $N(1.22, 0.08)$\\ [1ex] 
 \hline
 \end{tabular}
\end{minipage}
\caption[kepler]{(Left) Illustration of five of the eight Keplerian elements estimated in this work. The additional three elements not illustrated are the total mass of the exoplanet ($M_T$), the parallax ($\pi$), and the eccentricity ($e$). Eccentricity is defined by semi-major axis ($a$) and semi-minor axis ($b$) as $e = \sqrt{1-\frac{b^2}{a^2}}$. (Right) Definitions of Keplerian elements and their prior distributions used in exoplanet $\beta \mbox{ Pic } b$ astrometry.}
\label{fig:kepler}
\end{figure}

Detection and characterization of exoplanets using direct imaging is one of the most exciting frontiers in astronomy in the era of 30 meter class optical/infra-red telescopes. 
By capturing a series of snapshot images of an exoplanet, we can understand the properties of its planetary system \citep{bate2010chaotic, maire2019hint, scharf2009long, yu2001resonant}. Fitting an exoplanet's astrometry data allows us to understand the planet's formation and evolution, and sometimes even detect unseen planets~\citep{lacour2021mass}.
A planet's orbit is parameterized by eight Keplerian elements \citep{blunt2017orbits, blunt2020orbitize}: semimajor axis ($a$), eccentricity ($e$), inclination angle ($i$), argument of periastron of the secondary's orbit ($\omega$), longitude of ascending node ($\Omega$), epoch of periastron passage ($\tau$ , defined as a fraction of the orbital period past a reference epoch), parallax ($\pi$), and total mass ($M_T$), as illustrated in Fig.~\ref{fig:kepler}. In an orbit fitting problem, we estimate the posterior of these parameters based on the exoplanet astrometry data relative to the primary star (e.g., right ascension and declination) from telescope snapshot images.



\begin{figure}[tb]
\centering
\setlength{\fboxrule}{0pt}
\includegraphics[width=0.99\textwidth]{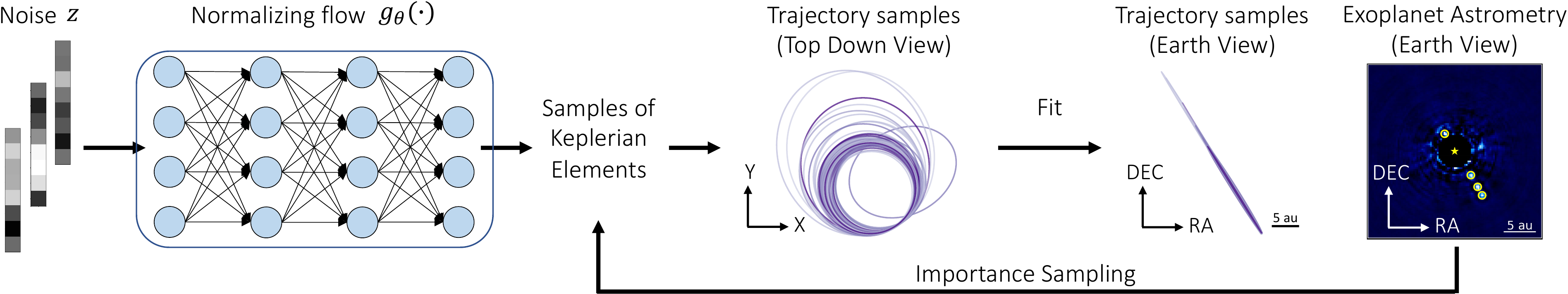}
\caption[planetinference]{$\alpha$-DPI applied to exoplanet astrometry. In this example, we infer the orbital parameters of an exoplanet, $\beta \mbox{ Pic } b$, based its snapshot images captured by the Gemini Planet Imager (GPI). $\alpha$-DPI first optimizes the normalizing flow weights, $\theta$, to generate diverse trajectory samples that match the observational data. Then it conducts importance sampling to re-weight each trajectory sample to produce an accurate posterior.In the plots, the trajectories with darker colors have higher posterior likelihoods. In the composite exoplanet astrometry image on the far right the location of the exoplanet at different times have been circled for clarity.  }
\label{fig:planetinference}
\end{figure}

\begin{figure}[!t]
\centering
\setlength{\fboxrule}{0pt}
\includegraphics[width=0.99\textwidth]{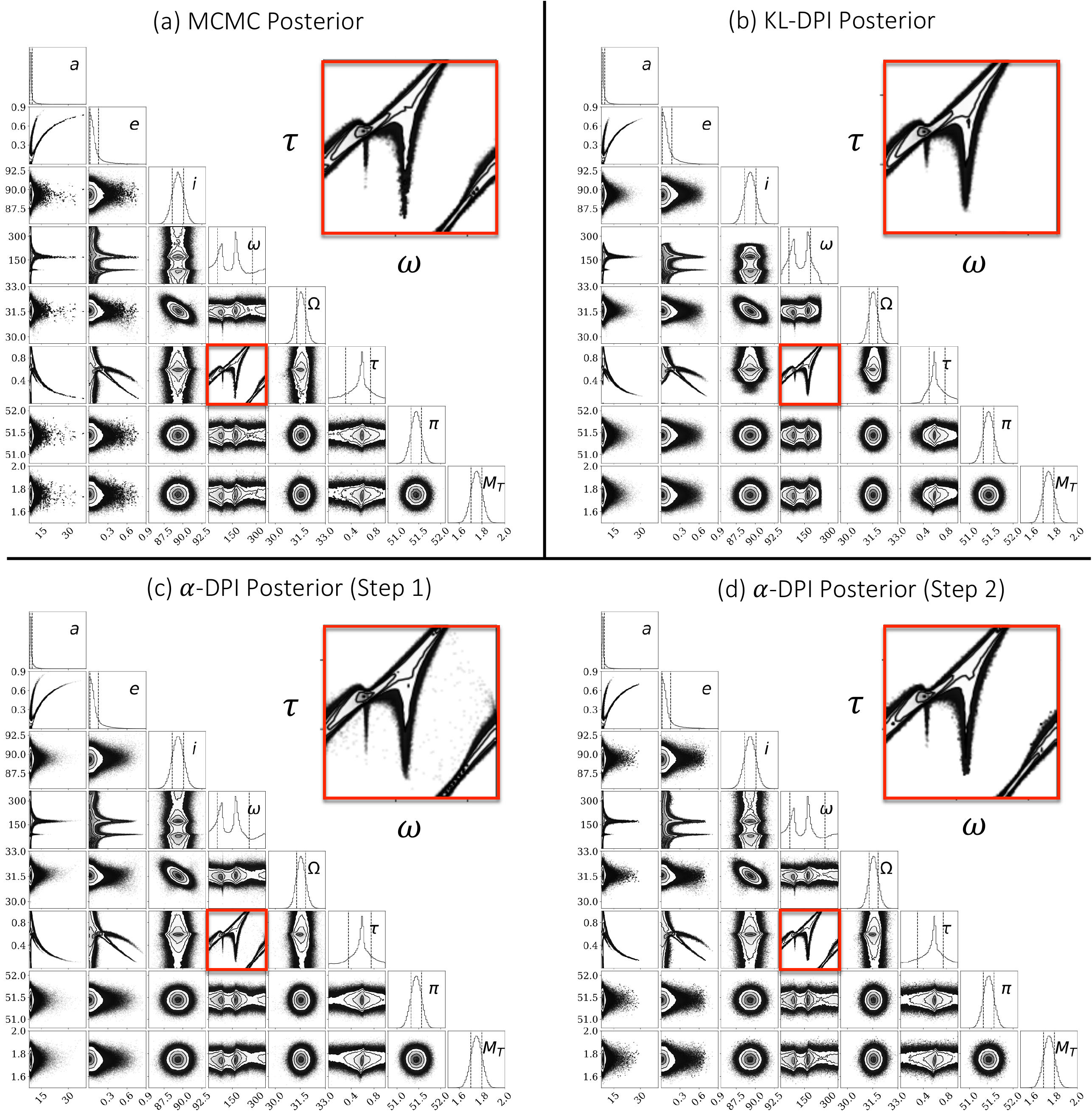}
\caption[orbitcorner]{Corner plots of posterior samples from (a) MCMC, (b) KL-DPI (equivalent to $\alpha$-DPI with $\alpha=1.0$, see Sec.~\ref{subsubsec:alphadiv}) and $\alpha$-DPI with $\alpha=0.5$ before (c; step 1) and after (d; step 2) importance sampling. 
The zoomed-in figures present the marginal joint distribution of argument of periastron ($\omega$) and epoch of periastron ($\tau$); the presence of disconnected modes is due to the periodicity of both parameters. KL-DPI collapses to a distribution that fails to capture some of the modes in the posterior. In contrast, $\alpha$-DPI captures all disconnected modes. Although (c) includes a few ``bad" samples that should have low probability, after importance sampling the resulting posterior in (d) is nearly identical to the MCMC identified posterior.
The contours in the corner plots represent the levels of the posterior likelihoods; points lying outside the posterior contours are the samples whose likelihoods are smaller than a threshold but not negligible. 
}
\label{fig:orbitcorner}
\end{figure}

We applied $\alpha$-DPI to a dataset from the Gemini Planet Imager (GPI) to infer the orbital parameters of a planet, $\beta$~Pictoris~b ($\beta \mbox{ Pic } b$) \citep{Wang2016}. The prior distributions on $\beta \mbox{ Pic } b$'s orbital parameters, as mentioned in \cite{Wang2016}, are listed in Fig.~\ref{fig:kepler}. As shown by the exoplanet astrometry data (right image) in Fig.~\ref{fig:planetinference}, $\beta \mbox{ Pic } b$ has an almost edge-on orbit when observed from the Earth. Due to the limited observational data, the posterior of $\beta \mbox{ Pic } b$'s Keplerian elements is multi-modal. 
Figure~\ref{fig:orbitcorner} shows the posterior samples from MCMC, KL-DPI, and $\alpha$-DPI before and after importance sampling (unless otherwise noted, $\alpha$-DPI includes the step of importance sampling). The posterior samples are visualized using corner plots, which present the marginal distribution of each Keplerian element and the joint distribution of each pair of Keplerian elements. As can be seen, KL-DPI neglects disconnected minor modes in its approximated posterior; this occurs because the KL loss strongly discourages including any mass from low posterior probability regions in $q_{\theta}(x)$, which is needed to connect disconnected modes in a posterior when using a normalizing flow model. 
For this reason, KL-DPI results in a distribution $q_{\theta}(x)$ that under-estimates the uncertainty of the planet's orbit.
In contrast, the normalizing flow learned in $\alpha$-DPI captures all the posterior modes of the astrometry data. Although $\alpha$-DPI includes several samples with low data likelihood that connect different posterior modes, it still provides a good proposal distribution that can be used by importance sampling to efficiently generate more accurate posterior samples. The final $\alpha$-DPI approximate posterior distribution achieves comparable accuracy to the MCMC sampler, but only takes 1.5 hours of computation on a single GTX1080 Ti GPU (time of importance sampling step is negligible compared with normalizing flow training). In comparison, the MCMC method requires more than 24 hours to recover the posterior \citep{blunt2020orbitize}. 

\begin{figure}[!t]
\centering
\setlength{\fboxrule}{0pt}
\includegraphics[width=0.9\textwidth]{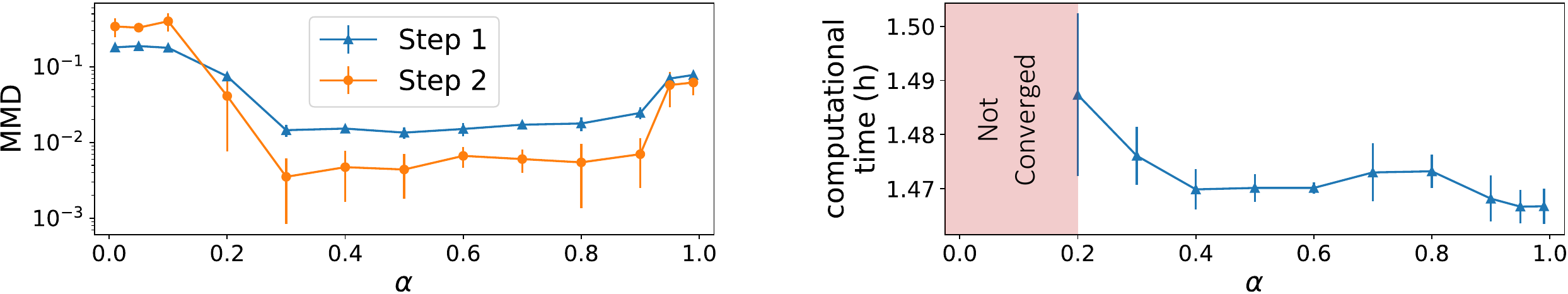}
\caption[orbitcurve]{\textcolor{black}{Posterior estimation precision (Left) and computational time (Right) for $\alpha$-DPI with different $\alpha$ values ($[0.01, 1]$). The error bars in both figures are generated based on 5 independent runs for each $\alpha$ value. The posterior estimation precision is evaluated using the maximum mean discrepancy (MMD) between the converged MCMC samples and $\alpha$-DPI samples, where smaller MMD indicates higher precision. Step 1 and Step 2 represent the samples before and after importance sampling, respectively. The computational time is defined as the time when the MMD of normalizing flow samples (Step 1) drops below $0.1$ for 30 consecutive iterations. The normalizing flow models optimized with $0<\alpha<0.2$ never achieve sufficient precision (MMD$<0.1$) in the allotted time ($\sim$ 3.5 hours), and thus are marked as ``not converged" even if they have converged to a distribution with a larger MMD. }   
}
\label{fig:orbitcurve}
\end{figure}

\textcolor{black}{We also quantitatively investigate the influence of the $\alpha$ parameter values on both $\alpha$-DPI's posterior estimation precision and computational time. The precision is evaluated using the maximum mean discrepancy (MMD) \citep{gretton2012kernel} between the ``ground-truth" MCMC posterior, $q_{mc}(\cdot)$, and the optimized normalizing flow posterior, $q_{dpi}(\cdot)$,
\begin{equation} \label{eq:mmd}
    \begin{split}
        MMD^2[q_{mc}(\cdot), q_{dpi}(\cdot)] = \mathbb{E}_{x, x' \sim q_{mc}(\cdot)}[K(x, x')] + \mathbb{E}_{y, y' \sim q_{dpi}(\cdot)}[K(y, y')] - 2 \mathbb{E}_{x \sim q_{mc}(\cdot), y \sim q_{dpi}(\cdot)}[K(x, y)],
    \end{split}
\end{equation}
where $K(\cdot, \cdot)$ is a kernel function between two samples. Smaller MMD indicates that two distributions are more similar to each other. A radial basis function (RBF) kernel \citep{scholkopf1997support} is used in our MMD calculation. The MMD of normalizing flow samples before and after importance sampling are both reported. As demonstrated in Fig.~\ref{fig:orbitcurve} (Left), the posterior estimation performs well for $\alpha$ between 0.3 and 0.9, but results in poor performance when $\alpha$ is either large or small. Large $\alpha$ leads to the mode collapse problem and additional importance sampling does not help since a posterior mode has already been neglected; small $\alpha$ makes the stochastic loss in Eq.~\ref{eq:method-vi} hard to evaluate with a limited batch size, resulting in a normalizing flow model that fails to converge to an informative proposal posterior. The computational time of $\alpha$-DPI is defined as the training time for the normalizing flow model to reach a sufficient precision (MMD$<0.1$) and stay at this precision for more than 30 iterations. We find the time spent on importance sampling step (Step 2) is negligible compared with the neural network training, even in the case of low $\alpha$. As shown in Figure~\ref{fig:orbitcurve} (Right), the computational time does not change significantly given a fixed normalizing flow architecture and a fixed batch size. The computational time using a smaller $\alpha$ is a little longer but varies more among different runs. This again indicates the evaluation of stochastic loss in Eq.~\ref{eq:method-vi} is difficult with small $\alpha$.}

Given an appropriate $\alpha$, the results of $\alpha$-DPI agree well with the posteriors from \citet{Wang2016}, indicating an accurate estimation of the possible orbital configurations for $\beta \mbox{ Pic } b$, including the fact it will not transit the star. However, due to its much faster run-time, $\alpha$-DPI lowers the barrier for obtaining orbital posteriors for a large number of systems in the future that have yet to be analyzed fully. For example, \citet{FerrerChavez2021} was able to explore biases in orbit fitting for short orbital arcs (spanning $\sim$1\% of a planet's orbit), but chose not to explore longer orbital arcs because current orbit fitting algorithms were prohibitively slow. Thus, $\alpha$-DPI holds significant promise for handling this inference problem both efficiently and accurately.

\paragraph{Implementation Details} In order to solve for $g_{\theta}(\cdot)$ using gradient-based optimization, the forward model that maps Keplerian elements to astrometry data is implemented in a differentiable manner. In particular, Kepler's equation is approximated via a Gauss-Newton solver with 10 fixed gradient descent optimization steps. We compare $\alpha$-DPI ($\alpha=0.5$ in the divergence loss) to a normalizing flow based KL-divergence VI method (denoted as KL-DPI in this paper) and a parallel-tempered MCMC (PTMCMC) sampler from ptemcee \citep{vousden2016dynamic}
using 1000 Markov chains with 45,000 iterations in each chain (first 40,000 steps as the burn-in). Since the MCMC sampler uses a large number of chains and iterations, we believe it converges to close-to the ground truth posterior. Therefore, we use the resulting MCMC posterior in order to evaluate the inference accuracy of $\alpha$-DPI.

\subsection{Black hole feature extraction with the Event Horizon Telescope (EHT)} \label{subsec:eht}
Very long baseline interferometry (VLBI) has enabled the reconstruction of high-resolution astronomical images using sparse data from a multi-telescope synthetic aperture. By joining radio telescopes from across the globe, the Event Horizon Telescope (EHT) collaboration captured the first picture of a black hole, M87$^*$ \citep{M87PaperIV}. This marked a new era for black hole astronomy, since scientists can now study black holes by directly observing their event-horizon scale structure. However, VLBI data for black hole imaging is typically very sparse and noisy: only seven telescopes at five geographic sites were used for collecting M87$^*$ data that led to the first black hole image -- this sparse telescope data was heavily corrupted by atmospheric turbulence and instrumentation calibration errors. In order to deliver reliable scientific interpretations, it is important to carefully characterize the uncertainty in features of the black hole image, including diameter, width, asymmetry, and position angle.

\begin{figure}[tb]
\centering
\setlength{\fboxrule}{0pt}
\includegraphics[width=0.99\textwidth]{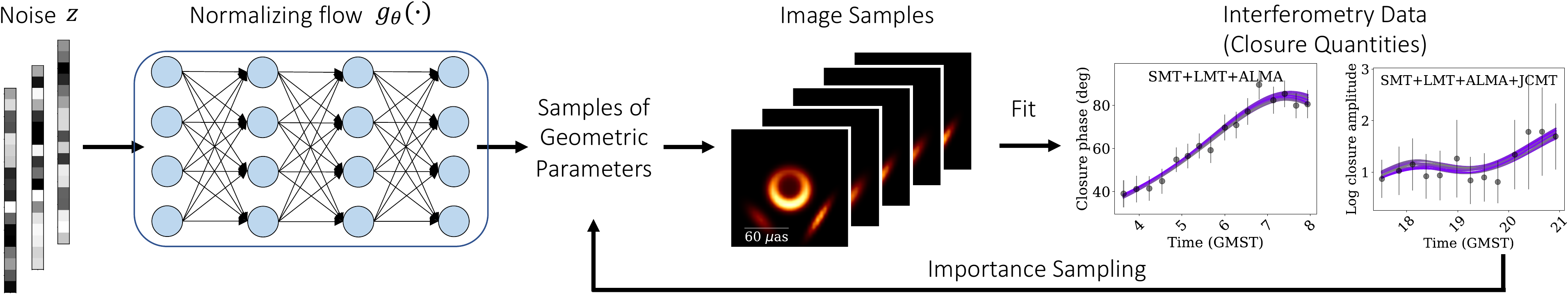}
\caption[vlbiinference]{$\alpha$-DPI applied to VLBI black hole feature extraction. In this example, we infer the geometric parameters (e.g., diameter, position angle, central emission) of a super massive black hole based on the data captured by the Event Horizon Telescope (EHT). We conduct experiments using both simulated interferometric data and real data of M87$^*$ captured in 2017. Due to atmospheric turbulence and instrument calibration errors, we fit robust data products referred to as closure quantities. The right two figures show the measurement data (gray dots with error bars) and the ideal corresponding measurements from posterior samples (purple curves; darker curves indicating more likely samples) of two closure quantities used to define the data likelihood. Here the capitalized letters within each plot represent the telescopes used for capturing the corresponding measurements.
}
\label{fig:vlbiinference}
\end{figure}

\begin{figure}[!h]
\begin{minipage}[t]{0.3\linewidth}
\vspace{2pt}
\centering
\includegraphics[width=0.99\textwidth]{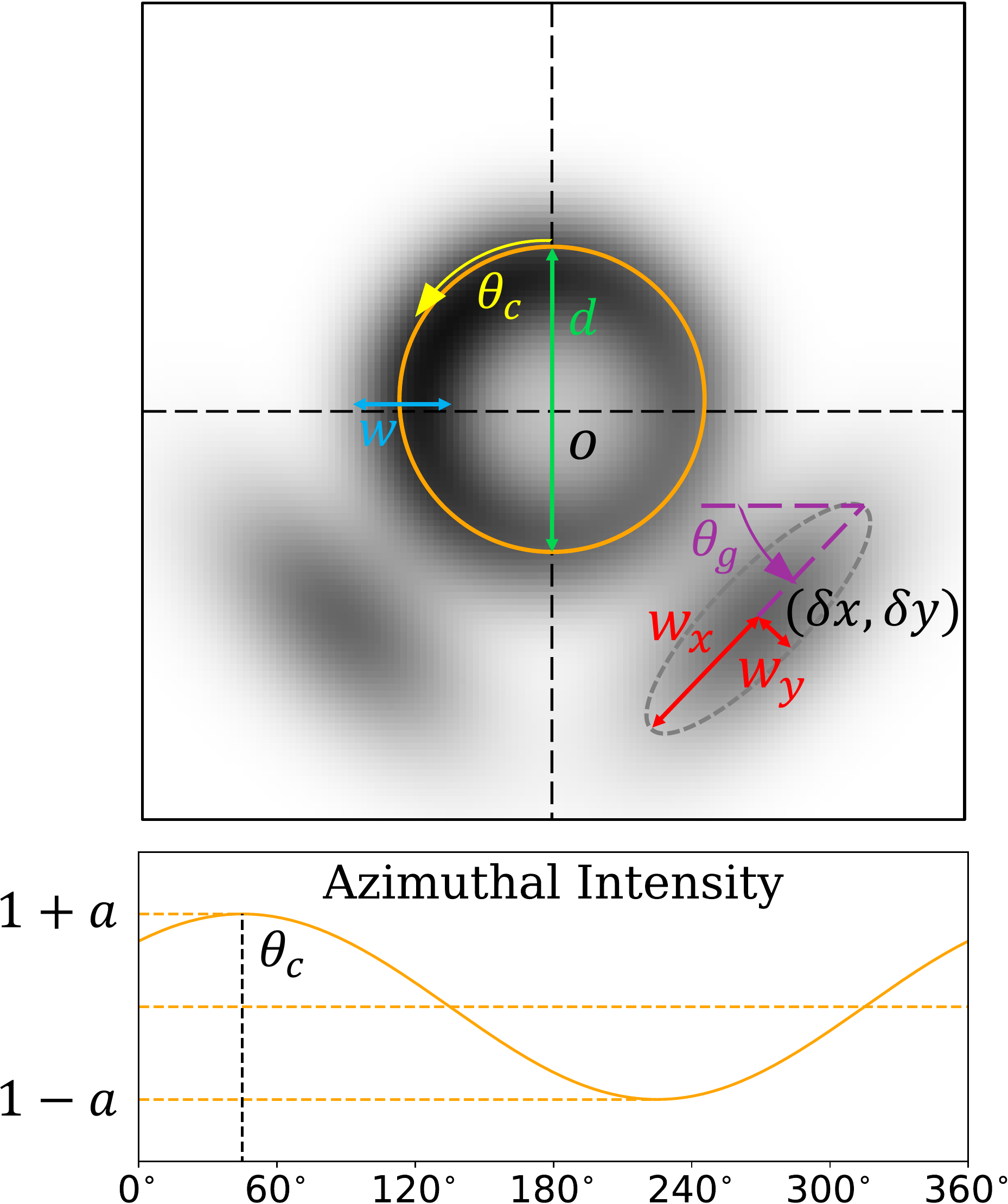}
\end{minipage}\hfill
\begin{minipage}[t]{0.7\linewidth}
\vspace{0pt}
\centering
  \begin{tabular}{c c c} 
 \hline
 Parameter & Unit & Prior distribution\\ [0.5ex] 
 \hline\hline
crescent diameter ($d$) & \vtop{\hbox{microarcsecond}\hbox{\quad \quad \ ($\mu as$)}} & uniform: $U(20, 100)$\\ 
 crescent width (fwhm, $w$) & \vtop{\hbox{microarcsecond}\hbox{\quad \quad \ ($\mu as$)}} & uniform: $U(1, 40)$\\ 
 crescent asymmetry ($a$) & $-$ & uniform: $U(0, 1)$\\ 
 crescent position angle ($\theta_c$)  & degree ($\circ$) & uniform: $U(0, 360)$\\ 
 crecent flux ($V_c$) & $-$ & uniform: $U(0, 2)$\\ \vtop{\hbox{flux ratio of central emission}\hbox{\quad \ disk and crescent ($V_d$)}} & $-$ & uniform: $U(0, 1)$\\ 
 \vtop{\hbox{central coordinates of additional}\hbox{Gaussian components ($\delta x_k$, $\delta y_k$)}}
 & \vtop{\hbox{microarcsecond}\hbox{\quad \quad \ ($\mu as$)}} & uniform: $U(-200, 200)$\\ 
 \vtop{\hbox{half major and minor axes of additional}\hbox{\ Gaussian components ($w_{x, k}$, $w_{y, k}$)}} & \vtop{\hbox{microarcsecond}\hbox{\quad \quad \ ($\mu as$)}} & uniform: $U(0, 100)$\\ 
 Gaussian component's position angle ($\theta_{g, k}$)  & degree ($\circ$) & uniform: $U(0, 90)$\\ 
 Gaussian component's flux ($V_{g, k}$) & $-$ & uniform: $U(0, 2)$\\ [1ex] 
 \hline
 \end{tabular}
\end{minipage}
\caption[crescent]{(Left) Illustration of a black hole geometric model. The model consists of two parts: 1) an asymmetric ring with a central emission disk (ring and disk share the same diameter), and 2) multiple Gaussian ellipses (index by $k$). The azimuthal intensity of the ring follows a sinusoidal function. The flux of the ring, central disk and Gaussian ellipses are $V_c$, $V_d$ and $V_{g, k}$, respectively, which are neglected in the figure. (Right) Definitions of black hole geometric parameters and their prior distributions used for simulated and real M87$^*$ interferometry data.}
\label{fig:crescent}
\end{figure}


In this section, we apply our proposed $\alpha$-DPI algorithm to both simulated and real EHT M87$^*$ data, and fit geometric models to understand important black hole properties, as shown in Fig.~\ref{fig:vlbiinference}. 
 We choose to paramterize the black hole image as the summation of a crescent (an asymmetric ring with a central emission disk) and multiple additional elliptical Gaussians, as illustrated in Fig.~\ref{fig:crescent}. Elliptical Gaussians are introduced to account for extended flux in the black hole image. The number of ellipses to include is unknown {\it a priori} when analyzing real observational data. Therefore, we first attempt models with different number of Gaussian ellipses (0-3) in our following experiments, and subsequently select the best geometric model by comparing the evidence lower bounds (ELBO) of different models, denoted as $m$. In particular, by expanding out the equation of $p(y | m)$ (i.e., probability of data under a given model) one can see that, if the true posterior has been well approximated (i.e., $D_{KL} \approx 0$), the ELBO function identifies which model best describes the data without over-interpretation: 
\begin{align}
\label{eq:elbo}
\begin{split}
\log p(y | m ) &= D_{KL}[q_{\theta}(x)\|p(x|y, m)] - \mathbb{E}_{x \sim q_{\theta}(x)}[- \log p(y|x, m) - \log p(x|m) + \log q_{\theta}(x)] \\
&\geq - \mathbb{E}_{x \sim q_{\theta}(x)}[- \log p(y|x, m) - \log p(x|m) + \log q_{\theta}(x)] 
\\
&= \text{ELBO}(m)
\end{split}
\end{align}
 Since $\alpha$-DPI samples often approximate the true posterior well, we assume that $D_{KL}$ 
 is negligible compared with the ELBO and choose the preferred model, $m$ by selecting the model with the largest ELBO value \citep{penny2012comparing}.

\begin{figure}[!h]
\begin{minipage}[t]{0.62\linewidth}
\vspace{0pt}
\centering
\includegraphics[width=1.00\textwidth]{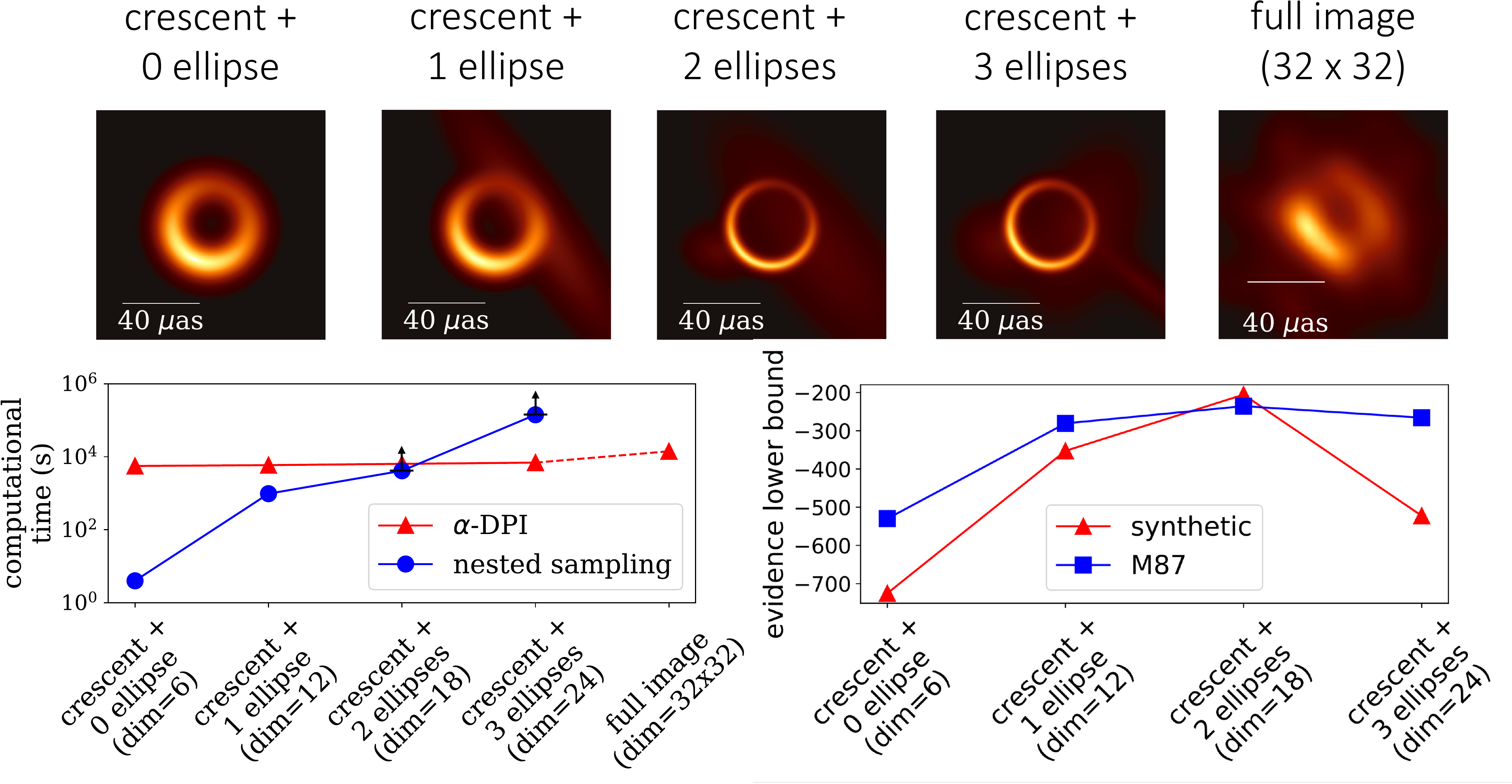}
\end{minipage}\hfill
\begin{minipage}[t]{0.38\linewidth}
\vspace{0pt}
\textbf{Comparison between the $\alpha$-DPI estimated M87$^{\star}$ parameters (2 Gaussian ellipses) and the original M87$^{\star}$ feature extraction results \citep{M87PaperVI}.}
\centering
 \begin{tabular}{c c c c} 
 \hline
 \texttt{Code} & \texttt{THEMIS} & \texttt{dynesty} & $\alpha$\texttt{-DPI} \\ [0.5ex] 
 \hline\hline
 $d$ ($\mu as$) & $43.5_{-0.14}^{+0.14}$ & $43.4_{-0.26}^{+0.27}$ & $43.2_{-0.25}^{+0.32}$ \\
 $\log_{10} (\hat{s})$ & $-0.87_{-0.20}^{+0.09}$ & $-1.31_{-0.52}^{+0.31}$ & $-1.00_{-0.17}^{+0.12}$ \\
 $\log_{10} (V_d)$ & $-2.14_{-0.62}^{+0.43}$ & $-2.63_{-0.60}^{+0.41}$ & $-1.94_{-0.62}^{+0.40}$ \\
 $a$ & $-$ & $-$ & $0.74_{-0.032}^{+0.031}$ \\
 $\theta_c$ (${\circ}$) & $153.0_{-2.4}^{+2.0}$ & $148.5_{-1.2}^{+1.4}$ & $153.1_{-1.6}^{+1.5}$\\ [1ex] 
 \hline
 \end{tabular}
 \\ $^{\star}$ $\hat{s}= \frac{2 \sigma \sqrt{2\ln{2}}}{d}$ is the crescent's sharpness.
\end{minipage}\vfill
\begin{minipage}[t]{1.00\linewidth}
\vspace{20pt}
\centering
\centering
\setlength{\fboxrule}{0pt}
\includegraphics[width=0.99\textwidth]{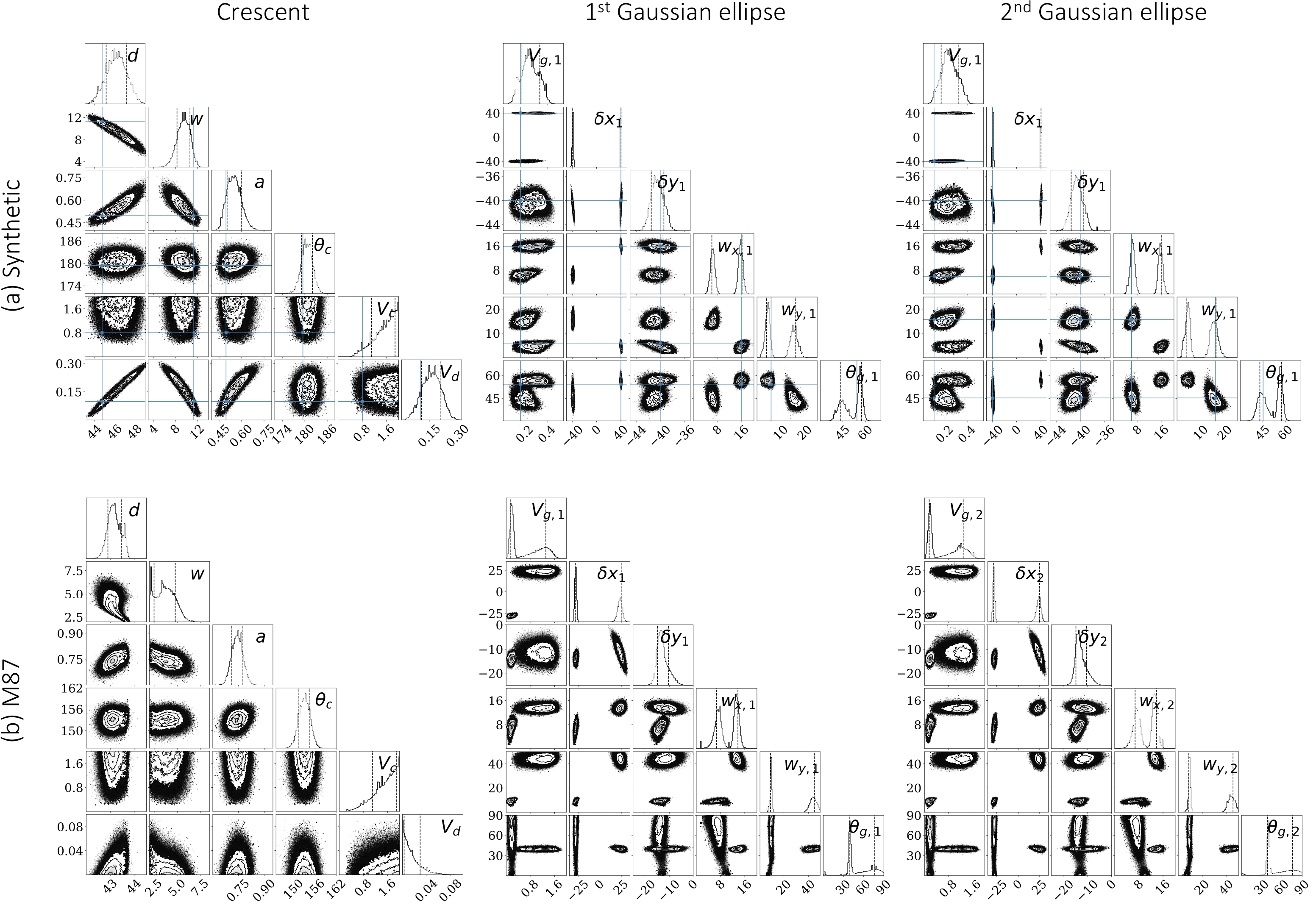}
\end{minipage}
\caption[vlbicorner]{(Top) Computational times and ELBO values obtained with models of increasing complexity. The computational time of nested sampling grows more than polynomially as the geometric model dimension increases. In contrast, the computational time of $\alpha$-DPI stays within the same order of magnitude.
Blue arrows on the sampling method's computational time indicate that there is evidence that this method has not fully converged in the allotted time. 
A crescent with two elliptical Gaussian ellipses is identified as the best model for both synthetic and real M87$^*$ data according to the ELBO. This choice correctly identifies the true underlying model used for generated synthetic data. It also agrees with the model choice selected and presented in~\cite{M87PaperVI} for the real M87$^*$ data (top right table, \texttt{THEMIS} and \texttt{dynesty} are two benchmark EHT feature extraction pipelines). (Bottom) Corner plots visualizing posterior samples of black hole geometric parameters (a crescent with two elliptical Gaussian ellipses). (a) and (b) figures are the results of synthetic and real M87$^*$ data, respectively. In the synthetic data experiment, the ground truth values (blue lines) are contained in the recovered posterior. Since the two Gaussian ellipses are interchangeable, in both synthetic and real data experiment $\alpha$-DPI correctly identifies that the marginal posteriors of two Gaussian ellipses are identical. The contours in the corner plots represent the levels of the posterior likelihoods; points lying outside the posterior contours are the samples whose likelihoods are smaller than a threshold but not negligible. }
\label{fig:vlbicorner}
\end{figure}

Simulated EHT data is generated using a crescent model with two elliptical Gaussians. Figure~\ref{fig:vlbicorner} (Top) presents the ELBO when fitting simulated data using models of varying complexity. As expected, the true model -- a crescent with two Gaussian ellipses -- has the largest ELBO value. The bottom figure in Fig.~\ref{fig:vlbicorner} (Bottom; a) shows the corner plots of $\alpha$-DPI posterior samples assuming a two-ellipse crescent model (cross correlation between the crescent and ellipses are neglected for brevity). The ground-truth values (blue lines) successfully lie in the recovered posterior. More interestingly, since the parameters of two Gaussian ellipses are interchangeable, their marginal posteriors are identical to each other; $\alpha$-DPI successfully identifies this predictable symmetry. 

We also investigate the posterior distribution of M87$^{*}$ geometric features using the real EHT dataset \citep{M87PaperII} that produced the first black hole image in 2019\footnote{This dataset is publicly available online at \url{https://eventhorizontelescope.org/for-astronomers/data}}.
The top left of Fig.~\ref{fig:vlbicorner} presents the mean reconstructed images obtained from $\alpha$-DPI samples under models of increasing complexity, along with the the corresponding computational time and ELBO values. 
Fig.~\ref{fig:vlbicorner} also presents the uncertainty of estimated black hole feature parameters, visualized by the corner plots (b) of $\alpha$-DPI samples using the model with the largest ELBO (a crescent with two ellipses). As seen in the corresponding table on the top right of Fig.~\ref{fig:vlbicorner}, the recovered posterior of black hole features aligns well with the sampling-based results first presented in \cite{M87PaperVI}. 

In addition to the inference accuracy, we also quantitatively study $\alpha$-DPI's efficiency by comparing its computational time to a representative nested sampling method \citep{speagle2020dynesty}. \textcolor{black}{In this section, the computational time is defined as the training time for $\alpha$-DPI to converge to the optimal loss or for nested sampling to reach the stationary state.} 
Fig.~\ref{fig:vlbicorner} presents results obtained using models with increasing complexity, ranging from 6 to 1024 dimensions. As shown in the top left of Fig.~\ref{fig:vlbicorner}, $\alpha$-DPI scales very well with the dimension of the model, remaining within the same order of magnitude for computational time across all models (refer to the computational time log-plot of Fig.~\ref{fig:vlbicorner}). \textcolor{black}{As $\alpha$-DPI is primarily an optimization-based problem, its computational time directly depends on the normalizing flow network architecture (depth, width, activation function, etc.) instead of the model dimension. As introduced in Sec.~\ref{subsec:implementation}, our experiments only need to use neural network architectures with the same depth (number of affine coupling layers) but varying widths (number of neurons in each layer) to handle geometric models with different dimensions. $\alpha$-DPI's computational time only slightly changes as the neural network width proportionally increases with the model dimension since it can make use of GPU parallel acceleration. As a comparison, nested sampling cannot be easily parallelized, so its computational time increases significantly with the model dimension. \citep{salomone2018unbiased}}
Note that since $\alpha$-DPI is much more computationally efficient than the baseline sampling methods, it easily scales up to estimating posteriors of full images with $32\times 32 = 1024$ parameters (i.e., pixels), which was demonstrated in \cite{sun2021deep} using KL-DPI. However, such a high-dimensional full image uncertainty quantification task is not easily achievable using current sampling-based methods, and thus for computational reasons previous methods are restricted to characterizing images with far fewer pixels \citep{broderick2020hybrid}.
$\alpha$-DPI's efficiency will become increasingly critical in future observations using the next-generation EHT (ngEHT) where we would like to fit higher-dimensional geometric models or images to understand higher-resolution features of black holes \citep{raymond2021evaluation}. 

\paragraph{Implementation Details} The observational data in VLBI black hole observations are the Fourier components of the astronomical signals, which are referring to ``visibilities" in radio astronomy \citep{thompson2017interferometry}. In practical observations, the visibilities are usually contaminated by atmospheric turbulence, so we define the $\alpha$-DPI target posterior using two robust data products derived from visibilities: closure phase and closure amplitude \citep{thompson2017interferometry, chael2018interferometric}. Since the exact likelihood of these closure quantities cannot be explicitly defined, we use Gaussian distributions to approximate their likelihood in our implementation.  This Gaussian approximation is reasonable with high SNR measurement data. We use $\alpha=0.9$ in the $\alpha$-DPI divergence loss for the black hole imaging; this $\alpha$ value is empirically chosen by finding the lowest $\alpha$ where fewer than 50 percent of unique $\alpha$-DPI samples are rejected by importance re-sampling. 
In the $32 \times 32$ image posterior estimation, rather than using the prior distribution defined for the geometric model in Figure~\ref{fig:crescent}, we introduce an image prior/regularizer that combines maximum entropy \citep[MEM,][]{skilling1984maximum} and total squared variation \citep[TSV,][]{bouman1993generalized, kuramochi2018superresolution} regularizers; since the observational data is sparse, adding an image regularizer is necessary to recover an interpretable posterior that highlights black hole images with desirable low-level image statistics.

\section{Discussion} \label{sec:discussion}
$\alpha$-DPI provides a general framework for uncertainty quantification in Bayesian inference problems. It is broadly applicable to inference problems in astronomical science: the only requirement is a differentiable forward model, which is usually available (e.g., black hole interferometry) or can be approximated using numerical methods (e.g., exoplanet astrometry). $\alpha$-DPI merges the strengths of both sampling-based and optimization-based inference methods, resulting in a very efficient and scalable approach that still maintains posterior estimation accuracy. By pairing a normalizing  generative model along with an $\alpha$-divergence VI loss, $\alpha$-DPI is less frequently plagued by the mode collapse failures than KL-DPI. 
All the above strengths make $\alpha$-DPI a promising new method in not only astronomy, but also many other high-dimensional scientific inference problems, such as seismic tomography \citep[]{gao2021deepgem}, inverse material design \cite[]{sanchez2018inverse}, and biomedical imaging \cite[]{sun2021deep}.

\section{Method Background \& Details} \label{sec:methods}
In this section, we first describe in detail the two steps of $\alpha$-DPI: 1) $\alpha$-divergence variational inference ($\alpha$-VI) with a normalizing flow neural network (Sec.~\ref{subsec:vi}), and 2) importance sampling (Sec.~\ref{subsec:is}). Then we describe the implementation details (Sec.~\ref{subsec:implementation}), including neural network architectures and optimization choices made for the two astronomical applications studied in this work.

\subsection{$\alpha$-divergence Variational Inference ($\alpha$-VI) with Normalizing Flow} \label{subsec:vi}
VI is an approach that solves an optimization problem to estimate a posterior distribution. In VI, a family of density functions parameterized by $\theta$, $q_{\theta}(x)$, is defined. Then, optimization algorithms are used to find the parameters, $\theta^{\star}$, that best match the variational density function to the target posterior distribution, $p(x|y)$. The optimization problem of VI is
\begin{equation}
\label{eq:vi}
\theta^{\star} = \arg \min_{\theta} D[q_{\theta}(x)\|p(x|y)],
\end{equation}
where $D[\cdot \| \cdot]$ is a divergence function that measures the similarity between two distributions. Two major factors that influence the performance of VI are the objective divergence function and the modeling capacity of the variational density function.

\subsubsection{Renyi's $\alpha$-divergence} \label{subsubsec:alphadiv}
In traditional VI, Kullback-Leibler (KL) divergence \citep{kullback1951information} is typically used as the objective loss function:
\begin{equation}
\label{eq:klvi}
\begin{split}
    \theta^{\star} &= \arg \min_{\theta} D_{KL}[q_{\theta}(x)\|p(x|y)] \\
    &= \arg \min_{\theta} \mathbb{E}_{x \sim q_{\theta}(x)}[- \log p(x|y) + \log q_{\theta}(x)] \\
    &= \arg \min_{\theta} \mathbb{E}_{x \sim q_{\theta}(x)}[- \log p(y|x) - \log p(x) + \log q_{\theta}(x)].
\end{split}
\end{equation}
This KL loss is equivalent to jointly optimizing the maximum {\it a posteriori} (MAP) loss (first two terms) along with the entropy of the variational distribution (last term). 
Although KL-VI theoretically should produce an optimized density function that well matches the true posterior, in practice KL-VI usually leads to zero-forcing effects. In other words, the variational model tries to avoid including samples from low probability regions, often resulting in under-estimating the posterior and thus the uncertainty of inferred parameters. Zero-forcing effects are especially detrimental when estimating posteriors with disconnected modes. 

In $\alpha$-DPI, we alternatively define our VI objective function using the Renyi's $\alpha$-divergence \citep{li2016r}:
\begin{equation}
    \label{eq:alphaviflow}
    \begin{split}
        \theta^{\star} &= \arg \min_{\theta} D_{\alpha}[q_{\theta}(x)\|p(x|y)] \\
        &= \arg \min_{\theta} \frac{1}{\alpha-1} \log \mathbb{E}_{x \sim q_{\theta}(x)} \left[(\frac{p(x|y)}{q_{\theta}(x)})^{1-\alpha} \right] \\
        &= \arg \min_{\theta} \frac{1}{\alpha-1} \log \mathbb{E}_{x \sim q_{\theta}(x)} \{\exp [(1-\alpha) (\log p(y|x) + \log p(x) - \log q_{\theta}(x))]\}.
    \end{split}
\end{equation}
Renyi's $\alpha$-divergence is a more general class of similarity metrics between two distributions $q_{\theta}(x)$ and $p(x|y)$ -- when $\alpha \to 1$  $\alpha$-divergence converges to a KL-divergence, $D_{KL}[q_{\theta}(x)\|p(x|y)]$, making $\alpha$-VI same as KL-VI, which only exploits samples from $q_{\theta}(x)$ to compute distributions’ similarity; when $\alpha \to 0$  $\alpha$-divergence converges to $-\log \int_{q_{\theta}(x)>0} p(x)dx$, making $\alpha$-VI same as a MLE of $\theta$\citep{amari2001information}, which needs to explore the full probability space of $p(x)$. The former case is very efficient in computation but sometimes lacking posterior estimation accuracy, while the latter one produces a very accurate posterior but is computationally slow. By tuning the $\alpha$ value in $D_{\alpha}[q_{\theta}(x)\|p(x|y)]$, we balance the exploitation and the exploration in our posterior estimation, so $\alpha$-DPI can be both efficient and accurate.
In contrast with KL-VI, $\alpha$-VI ($0<\alpha<1$) can tolerate a few samples with low likelihood if it leads to a distribution that better captures multiple modes in the true posterior. 
Please refer to Appendix~\ref{appendix:klalpha} for more detailed derivations. 

\subsubsection{Normalizing Flow} \label{subsubsec:flow}
In order to solve the stochastic optimization problem in Eq.~\ref{eq:alphaviflow} efficiently, 
the variational density function should be efficient in both sampling ($x \sim q_{\theta}(x)$) and evaluating the likelihood of a sample (i.e., computing $\log q_{\theta}(x)$). Traditional VI typically uses a simple family of density functions, such as those from the exponential family, to facilitate efficient sampling and evaluation. However, these simple variational models sacrifice the ability to capture complex multi-modal distributions. 
To improve inference accuracy, in $\alpha$-DPI we propose to use a more flexible neural network based density function: a normalizing flow network. 
Normalizing flows are a class of deep generative models that are widely used in computer vision and machine learning for complex density function approximation. They parameterize a probability density $q_{\theta}(\cdot)$ in an implicit manner by transforming a simple base distribution $\pi(\cdot)$ using an invertible neural network,
\begin{equation}
\label{eq:INN}
    x = g_{\theta}(z), \quad z = g^{-1}_{\theta}(x),
\end{equation}
where $z \sim \pi(z)$ is an arbitrary sample from the base distribution (e.g., i.i.d. Gaussian), and $x \sim q_{\theta}(x)$ is a sample from the approximated distribution $q_{\theta}(x)$. 
According to the "change of variables theorem",
\begin{equation}
\label{eq:changeofvariable}
    q_{\theta}(x) = \pi(z) \left| \det \frac{d g_{\theta}(z)}{d z} \right|^{-1}
    = \pi(g_{\theta}^{-1}(x)) \left| \det \frac{d g^{-1}_{\theta}(x)}{d x} \right|,
\end{equation}
where $\det \frac{d g^{-1}_{\theta}(x)}{d x}$ is the determinant of the neural network function's Jacobian matrix. Therefore, the probability of a random sample from $q_{\theta}(x)$ is easy to evaluate when the Jacobian matrix is computationally tractable, as is the case with commonly used normalizing flow forms includes NICE \citep{dinh2014nice}, Real-NVP \citep{dinh2016density}, and Glow \citep{kingma2018glow}. As described in Section~\ref{subsec:implementation}, in this work we choose to use a Real-NVP generative network to parameterize $q_{\theta}(x)$.

\subsection{Importance Sampling}
\label{subsec:is}
Importance sampling is a Monte Carlo technique for estimating posterior distributions.  It  first generates random samples based on a proposal distribution $q(x)$, and then re-weights the samples based on their posterior probabilities $p(x|y)$, where the weight of each sample $x_j$ is $w_j = \frac{p(x_j|y)}{q(x_j)} \propto \frac{p(y|x_j)p(x_j)}{q(x_j)}$. In $\alpha$-DPI, importance sampling is applied after $\alpha$-VI to further improve the accuracy of the normalizing flow approximated posterior.

A normalizing flow, by definition, is a continuous bijective function. As a result, to capture multiple disconnected modes in a posterior distribution, a normalzing flow model must include low-probability samples that connect the modes. Therefore, to approximate the posterior more accurately, these samples should be removed from the approximated distribution. Since the learned normalizing flow distribution $q_{\theta}(x)$ is already close to the true posterior after $\alpha$-VI, $w_j = \frac{p(x_j|y)}{q(x_j)} \approx 1$ for most samples $x_j \sim q_{\theta}(x)$. In this case, applying the importance re-weighting to samples from the normalizing flow distribution is efficient because very few samples are rejected. The computational time of importance sampling is negligible compared with $\alpha$-VI, however, it produces a cleaner posterior estimation, as shown in Fig.~\ref{fig:orbitcorner}.

\subsection{Implementation Details} \label{subsec:implementation}
In both exoplanet astrometry and black hole feature extraction problems, we use a Real-NVP model with 32 affine coupling layers as the variational density function \citep{dinh2016density}. Each affine-coupling layer is composed of a neural neural network with 3 fully connected layers, where the width (number of neurons) of each fully connected layer is 16 times of  the dimension of inferred parameters.

The VI optimization problem is solved using an Adam optimizer \citep{kingma2014adam}. In addition, we apply simulated annealing training \citep{huang2018improving} to avoid the normalizing flow from stopping exploration in early epochs, and subsequently converging to a poor local minimum. Instead of directly optimizing the $\alpha$-divergence in Eq.~\ref{eq:alphaviflow}, we define an annealed objective function,
\begin{equation}
\label{eq:annealobj}
    D_{\alpha}^i[q_{\theta}(x)\|p(x|y)]= \arg \min_{\theta} \frac{1}{\alpha-1} \log \mathbb{E}_{x \sim q_{\theta}(x)} \{\exp [(1-\alpha) (\frac{1}{\beta_i}\log p(y|x) + \frac{1}{\beta_i} \log p(x) - \log q_{\theta}(x))]\}.
\end{equation}
Where $i$ is the index of the optimization epoch, $\beta_i = \max \{1, \beta_0 \exp(- \frac{i}{\tau})\}$, $\beta_0$ is the initial annealing weight, and $\tau$ is the decay period. Since the normalizing flow network used is initialized to an approximately random uniform distribution, the data likelihood ($\log p(y|x)$) and the prior likelihood ($\log p(x)$) are typically much larger than the entropy term ($\log q_{\theta}(x)$) at the initiation of training. $\beta_j$ balances the values of these different terms so that the optimization becomes more numerically stable and the normalizing flow can gradually converge from a random uniform distribution to one that well approximates the posterior. In exoplanet astrometry, we use an initial annealing weight $\beta_0 = 10^4$, a decay period $\tau=3000$ and in total $20,000$ training epochs. In black hole feature extraction, the initial annealing weight, the decay period, the total number of epochs are $\beta_0=10^3$, $\tau=3000$, and $epoch=15,000$, respectively.

\begin{acknowledgments}
This work was supported by NSF award 1935980, NSF award 2034306, NSF award 2048237, and Amazon AI4Science Fellowship. We thank the National Science Foundation (AST-1935980) for financial support of this work.
This work has been supported in part by the Black Hole Initiative at Harvard University, which is funded by grants from the John Templeton Foundation and the Gordon and Betty Moore Foundation to Harvard University. This work was supported in part by Perimeter Institute for Theoretical Physics.  Research at Perimeter Institute is supported by the Government of Canada through the Department of Innovation, Science and Economic Development Canada and by the Province of Ontario through the Ministry of Economic Development, Job Creation and Trade.  J.W. and S.B. are supported by the Heising-Simons Foundation (including grant 2019-1698). The authors would also like to thank Shiro Ikeda for the helpful discussions.
\end{acknowledgments}

%

\vspace{5mm}
\facilities{Gemini Planet Imager (GPI), Event Horizon Telescope (EHT)}


\software{astropy \citep{2013A&A...558A..33A,2018AJ....156..123A},  
          Cloudy \citep{2013RMxAA..49..137F}, 
          Source Extractor \citep{1996A&AS..117..393B}
          eht-imaging \citep{chael2018interferometric}
          }



\appendix
\section{More comparison of KL-DPI and $\alpha$-DPI} \label{appendix:klalpha}

Both KL-DPI and $\alpha$-DPI losses are approximated as stochastic forms when we solve for the normalizing flow weights:
\begin{equation}
    \label{eq:viflow}
    \begin{split}
    D_{KL}[q_{\theta}(x)\|p(x|y)] &= \mathbb{E}_{x \sim q_{\theta}(x)}[- \log p(y|x) - \log p(x) + \log q_{\theta}(x)] \\
    &\approx \sum_{n=1}^N [- \log p(y|g_{\theta}(z_n)) - \log p(g_{\theta}(z_n)) - \log \left| \det \frac{d g_{\theta}(z_n)}{d z_n} \right| + \log \pi(z_n)],\\
    D_{\alpha}[q_{\theta}(x)\|p(x|y)] \ \ 
    &= \frac{1}{\alpha-1} \log \mathbb{E}_{x \sim q_{\theta}(x)} \{\exp [(1-\alpha) (\log p(y|x) + \log p(x) - \log q_{\theta}(x))]\} \\
    &\approx \frac{1}{\alpha-1} \log \sum_{n=1}^N \exp \{(1-\alpha) (\log p(y|g_{\theta}(z_n)) + \log p(g_{\theta}(z_n)) + \log \left| \det \frac{d g_{\theta}(z_n)}{d z_n} \right|) - \log \pi(z_n)\}, \\
\end{split}
\end{equation}
where $z_n \sim \pi(\cdot)$ is an arbitrary sample from the base distribution, and $N$ is the number of samples used for Monte Carlo approximation. Denoting the KL loss function of a particular sample as $L_{\theta}(z_n) = - \log p(y|g_{\theta}(z_n)) - \log p(g_{\theta}(z_n)) - \log \left| \det \frac{d g_{\theta}(z_n)}{d z_n}\right| + \log \pi(z_n)$, the gradient of the KL-divergence and the $\alpha$-divergence are respectively
\begin{equation}
\label{eq:viflowgrad}
    \begin{split}
    \nabla D_{KL}[q_{\theta}(x)\|p(x|y)] &\approx \sum_{n=1}^N \nabla L_{\theta}(z_n)\\
    \nabla D_{\alpha}[q_{\theta}(x)\|p(x|y)] &\approx \frac{1}{\alpha-1} \nabla \log \{\sum_{n=1}^N \exp [- (1-\alpha) L_{\theta}(z_n)]\} \\
    &=\sum_{n=1}^N w_n \nabla L_{\theta}(z_n),
    \end{split}
\end{equation}
where $w_n = \text{Softmax}\{[- (1-\alpha) L_{\theta}(z_n)] \}$. The derivative of $\alpha$-divergence is a weighted version of the KL-divergence's derivative gradient such that the samples that lead to a large loss have singificantly less impact on gradient descent optimization in $\alpha$-DPI when compared to KL-DPI. As a result, $\alpha$-DPI allows the learned normalizing flow generative model to include a few ``bad" samples as long as the resulting distribution $q_{\theta}(x)$ better captures the modes in the true posterior distribution.

\bibliography{sample631}{}
\bibliographystyle{aasjournal}


\end{CJK*}
\end{document}